\documentclass[manuscript]{acmart}
\AtBeginDocument{%
  \providecommand\BibTeX{{%
    \normalfont B\kern-0.5em{\scshape i\kern-0.25em b}\kern-0.8em\TeX}}}

\copyrightyear{2022} 
\acmYear{2022} 
\setcopyright{acmcopyright}\acmConference[RecSys '22]{Sixteenth ACM Conference on Recommender Systems}{September 18--23, 2022}{Seattle, WA, USA}
\acmBooktitle{Sixteenth ACM Conference on Recommender Systems (RecSys '22), September 18--23, 2022, Seattle, WA, USA}
\acmPrice{15.00}
\acmDOI{10.1145/3523227.3546788}
\acmISBN{978-1-4503-9278-5/22/09}

\usepackage{subfig}
\usepackage{soul}
\usepackage{enumitem}
\usepackage{textcomp}
\usepackage{float}
\usepackage[linesnumbered,ruled]{algorithm2e}
\usepackage{siunitx}
\begin{document}
\fancyhead{}

\title{Denoising Self-Attentive Sequential Recommendation}

%\title{Prune Seq2Seq Model via Gradient Regularization}
%https://arxiv.org/pdf/2011.04419.pdf

%http://proceedings.mlr.press/v97/verma19a/verma19a.pdf
\author{Huiyuan Chen}
\email{hchen@visa.com}
\affiliation{%
  \institution{Visa Research}
  \country{USA}
}

\author{Yusan Lin}
\email{yusalin@visa.com}
\affiliation{%
  \institution{Visa Research}
  \country{USA}
}
\author{Menghai Pan}
\email{menpan@visa.com}
\affiliation{%
  \institution{Visa Research}
  \country{USA}
}

\author{Lan Wang}
\email{lwang4@visa.com}
\affiliation{%
  \institution{Visa Research}
  \country{USA}
}

\author{Chin-Chia Michael Yeh}
\email{miyeh@visa.com}
\affiliation{%
  \institution{Visa Research}
  \country{USA}
}

\author{Xiaoting Li}
\email{xiaotili@visa.com}
\affiliation{%
  \institution{Visa Research}
  \country{USA}
}

\author{Yan Zheng}
\email{yazheng@visa.com}
\affiliation{%
  \institution{Visa Research}
  \country{USA}
}

\author{Fei Wang}
\email{feiwang@visa.com}
\affiliation{%
  \institution{Visa Research}
  \country{USA}
}

\author{Hao Yang}
\email{haoyang@visa.com}
\affiliation{%
  \institution{Visa Research}
  \country{USA}
}

\begin{abstract}
Transformer-based sequential recommenders are very powerful for capturing both short-term and long-term sequential item dependencies. This is mainly  attributed to their unique self-attention networks to exploit pairwise item-item interactions within the sequence. However, real-world item sequences are often  noisy, which is particularly true for implicit feedback.  For example, a large portion of clicks do not align well with user preferences, and many products end up with negative reviews or being returned. As such, the current user action only depends on a subset of items, not on the  entire sequences.  Many  existing Transformer-based models use full attention distributions, which inevitably assign certain credits to irrelevant items. This may lead to sub-optimal performance if Transformers are not regularized properly. 

Here we propose the Rec-denoiser model for better training of self-attentive recommender systems. In  Rec-denoiser, we aim to adaptively prune noisy items that are unrelated to the next item prediction. To achieve this, we simply attach each self-attention layer with a trainable binary mask to prune noisy attentions, resulting in  sparse and clean attention distributions. This largely purifies item-item dependencies and provides better model interpretability.  In addition, the self-attention network is typically \textsl{not}  Lipschitz continuous and is  vulnerable to small perturbations. Jacobian regularization  is further applied to the Transformer blocks to improve the robustness of Transformers for noisy sequences.  Our Rec-denoiser is a general plugin that is compatible to many 
Transformers. Quantitative results on  real-world datasets show that our Rec-denoiser outperforms the state-of-the-art baselines.

\end{abstract}

%   <concept>
%       <concept_id>10002951.10003317.10003347.10003350</concept_id>
%       <concept_desc>Information systems~Recommender systems</concept_desc>
%       <concept_significance>500</concept_significance>
%       </concept>
%  </ccs2012>

% \ccsdesc[500]{Information systems~Recommender systems}
%\keywords{datasets, neural networks, gaze detection, text tagging}
\begin{CCSXML}
<ccs2012>
   <concept>
       <concept_id>10002951.10003317.10003347.10003350</concept_id>
       <concept_desc>Information systems~Recommender systems</concept_desc>
       <concept_significance>500</concept_significance>
       </concept>
 </ccs2012>
\end{CCSXML}

\ccsdesc[500]{Information systems~Recommender systems}
\keywords{Sequential Recommendation, Sparse Transformer, Noise Analysis, Differentiable Mask}

\maketitle

\section{Introduction}

Sequential recommendation  aims to recommend the next item based on a user's historical  actions~\cite{rendle2010factorizing,HidasiKBT15,ma2020disentangled,wang2021counterfactual,wang2021sequential},  \textsl{e.g.}, to recommend a bluetooth headphone after a user purchases a smart phone. Learning sequential user behaviors is, however, challenging since a user's choices on items generally  depend on both long-term and short-term preferences. Early Markov Chain models~\cite{rendle2010factorizing,he2017translation} have been proposed to capture short-term item transitions by assuming that a user's next decision is derived from a few preceding actions, while neglecting long-term preferences. To alleviate this limitation, many deep neural networks have been proposed to model the entire users' sequences and achieve great success, including recurrent neural networks~\cite{HidasiKBT15,xu2019recurrent} and convolutional neural networks~\cite{yan2019cosrec,yuan2019simple,tang2018personalized}.

Recently, Transformers have shown promising  results in various tasks, such as machine translation~\cite{vaswani2017attention}. One key component of Transformers is the self-attention network, which is capable of learning long-range dependencies by computing attention weights between each pair of objects in a sequence. Inspired by the success of Transformers, several self-attentive sequential recommenders have been proposed and achieve the state-of-the-art performance~\cite{kang2018self,sun2019bert4rec,wu2020sse,wu2020deja}. For example, SASRec~\cite{kang2018self} is the pioneering framework to adopt self-attention network to learn the importance of items at different positions. BERT4Rec~\cite{sun2019bert4rec} further models the correlations of items from both left-to-right and right-to-left directions. SSE-PT~\cite{wu2020sse} is a personalized Transformer model that provides better interpretability of engagement patterns by introducing user embeddings. LSAN~\cite{li2021lightweight} adopts a novel twin-attention sequential framework, which can capture both long-term and short-term user preference signals. Recently, Transformers4Rec~\cite{de2021transformers4rec} performs an empirical analysis with broad experiments of various Transformer architectures for
the task of sequential recommendation.

Although encouraging performance has been achieved,  the robustness of sequential recommenders is far less studied  in the literature. Many  real-world item sequences are naturally noisy, containing both true-positive and false-positive interactions~\cite{wang2021denoising,wang2021clicks,chen2021structured}.  For example, a large portion of clicks do not align well with user preferences, and many products end up with negative reviews or being returned.  In addition,  there is no any prior knowledge  about how a user's historical actions should be generated  in online systems. Therefore, developing robust algorithms to defend noise is of great significance for sequential recommendation.

Clearly, not every item in a sequence is aligned well with user preferences, especially for implicit feedbacks (\textsl{e.g.}, clicks, views, etc.)~\cite{chen2019behavior}.   Unfortunately, the vanilla self-attention network is \textsl{not}  Lipschitz continuous\footnote{Roughly speaking, a function is Lipschitz continuous
if changing its input by a certain amount will not significantly change its output.}, and is vulnerable to the quality of input sequences~\cite{kim2021lipschitz}.  Recently, in the tasks of language modeling, people  found that a large amount of BERT's attentions focus on less meaningful tokens, like "[SEP]" and ".", which leads to a misleading  explanation~\cite{clark2019does}.   It is thus likely to obtain sub-optimal performance if self-attention networks are not well regularized for noisy sequences. We use the following example to further explain above concerns.

\begin{figure}
	\begin{center}
	\includegraphics[width=6.0cm]{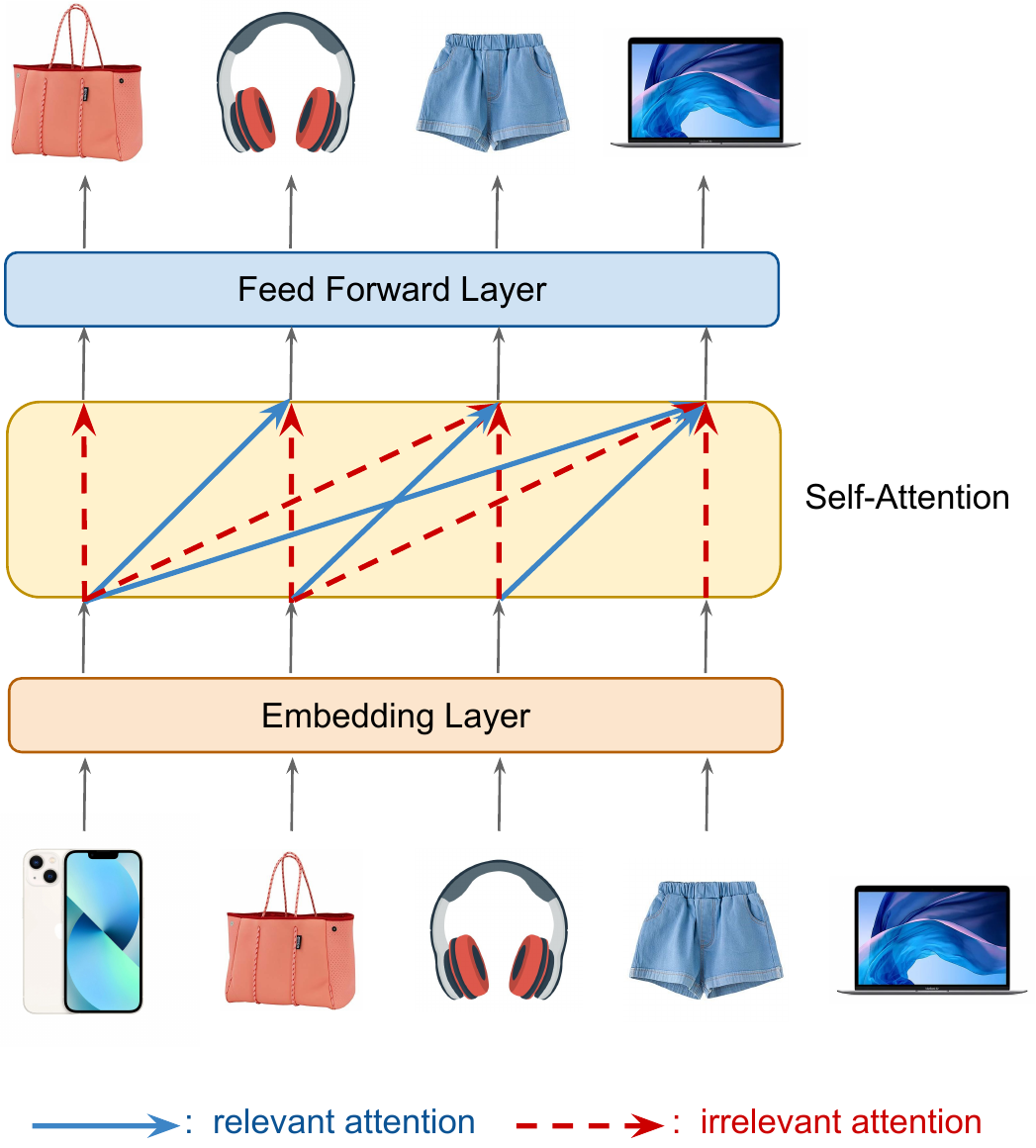}
	\end{center}
	\caption{An illustrative example of sequential recommendation where a sequence contains noisy or irrelevant items in left-to-right self-attention networks.}
	\label{graph}
\end{figure}

 Figure \ref{graph} illustrates an example of left-to-right sequential recommendation where a user's sequence contains some noisy or irrelevant items.   For example, a father may interchangeably purchase (\textsl{phone, headphone, laptop}) for his son, and  (\textsl{bag, pant}) for his daughter, resulting in a  sequence: (\textsl{phone, bag, headphone, pant, laptop}).  In the setting of sequential recommendation,  we intend to infer the next item, \textsl{e.g.}, \textsl{laptop}, based on the user's previous actions, \textsl{e.g.},  (\textsl{phone, bag, headphone, pant}). However, the correlations among items are unclear, and intuitively \textsl{pant} and \textsl{laptop} are neither complementary nor compatible to each other, which makes the prediction untrustworthy.  A trustworthy model should be able to only capture correlated items while ignoring these irrelevant items within sequences.  Existing self-attentive sequential models (\textsl{e.g.}, SASRec~\cite{kang2018self} and BERT4Rec~\cite{sun2019bert4rec}) are insufficient to address noisy items within sequences.   The reason is that their full attention distributions are \textsl{dense} and would assign certain credits to all items, including irrelevant items. This causes a lack of focus and makes models less interpretable~\cite{child2019generating,zaheer2020bigbird}. 

To address the above issues, one straightforward strategy  is to design \textsl{sparse} Transformer architectures that sparsify the connections in the attention layers, which have been actively investigated in language modeling tasks~\cite{child2019generating,zaheer2020bigbird}. Several representative models are Star Transformer~\cite{guo2019star}, Sparse Transformer~\cite{child2019generating}, Longformer~\cite{beltagy2020longformer}, and BigBird~\cite{zaheer2020bigbird}. These sparse attention patterns could mitigate noisy issues and avoid allocating credits to  unrelated contents for the query of interest.  However, these models largely rely on \textsl{pre-defined} attention schemas, which lacks flexibility and adaptability in practice. Unlike end-to-end training approaches, whether these sparse patterns could generalize well to sequential recommendation remains unknown and is still an open research question. \\

\noindent \textbf{Contributions.} In this work, we propose to design a denoising strategy, Rec-Denoiser, for better training of self-attentive sequential recommenders.  Our idea stems from the recent findings that  \textsl{not all} attentions are necessary and simply pruning redundant attentions could  further improve the performance~\cite{child2019generating,zaheer2020bigbird,correia2019adaptively,sukhbaatar2019adaptive,yeh2021embedding}. Rather than randomly dropping out attentions,  we introduce differentiable masks to drop task-irrelevant attentions in the self-attention layers, which can yield exactly zero attention scores for noisy items.

The introduced sparsity in the self-attention layers has several benefits: 1) Irrelevant attentions with parameterized masks can be learned to be dropped in a data-driven way. Taking Figure \ref{graph} as an example, our Rec-denoiser would prune the  sequence (\textsl{\st{phone}, {bag}, \st{headphone}}) for  \textsl{pant}, and (\textsl{phone, \st{bag}, headphone, \st{pant}}) for  \textsl{laptop} in the attention maps. Namely, we seek next item prediction explicitly based on a subset of more informative items.  2) Our Rec-Denoiser
still takes full advantage of Transformers as it does not change their architectures, but only the attention distributions. As such, Rec-Denoiser is easy to implement and is compatible to any Transformers, making them  less complicated as well as improving their interpretability. 

In our proposed Rec-Denoiser, there are two major challenges. First,  the discreteness of binary masks (\textsl{i.e.}, $0$ is dropped while $1$ is kept) is, however, intractable in the back-propagation. To remedy this issue, we relax the discrete variables with a continuous approximation through probabilistic reparameterization~\cite{jang2016categorical}.  As such, our differentiable masks can be trained jointly with original Transformers in an end-to-end fashion. In addition, the scaled dot-product attention  is \textsl{not}  Lipschitz continuous and is thus vulnerable to input perturbations~\cite{kim2021lipschitz}. In this work,   Jacobian regularization~\cite{jakubovitz2018improving,hoffman2019robust} is further applied to the entire Transformer blocks,  to improve the robustness of Transformers for noisy sequences. Experimental results on real-world benchmark datasets demonstrate the effectiveness and robustness of the proposed  Rec-Denoiser. In summary, our  contributions are:
    \begin{itemize}[leftmargin=*]
		\item We introduce the idea of denoising item sequences for better of training self-attentive sequential recommenders, which greatly reduces the negative impacts of noisy items.
		\item We present a general Rec-Denoiser framework with differentiable masks that can achieve sparse attentions by dynamically pruning irrelevant information, leading to better model performance. 
		\item We propose an unbiased gradient estimator to optimize the binary masks, and apply Jacobian regularization on the gradients of Transformer blocks to further improve its robustness.
		\item The experimental results demonstrate significant improvements that Rec-Denoiser brings to self-attentive recommenders ($5.05\% \sim 19.55\%$ performance gains), as well as its robustness against input perturbations.
	\end{itemize}

\section{Related Work}
 In this section, we briefly review the related work on sequential recommendation and sparse Transformers. We also highlight the differences between the existing efforts and ours.

	\subsection{Sequential Recommendation}
	 Leveraging sequences of user-item interactions is crucial for sequential recommendation.   User dynamics can be caught by Markov Chains for inferring the conditional probability of an item based on the previous items ~\cite{rendle2010factorizing,he2017translation}. More recently,  growing efforts have been dedicated to deploying deep neural networks for sequential recommendation such as recurrent neural networks~\cite{HidasiKBT15,xu2019recurrent}, convolutional neural networks~\cite{yan2019cosrec,yuan2019simple,tang2018personalized},  memory networks~\cite{chen2018sequential,huang2018improving}, and graph neural networks~\cite{wu2019session,chang2021sequential,chen2022graph}. For example, GRU4Rec~\cite{HidasiKBT15} employs a gated recurrent unit to study temporal behaviors of users. Caser~\cite{tang2018personalized} learns sequential patterns 
	 by using convolutional filters on local  sequences.  MANN~\cite{chen2018sequential} adopts memory-augmented neural networks to model user historical records. SR-GNN~\cite{wu2019session} converts  session sequences into graphs and uses graph neural networks to capture complex item-item transitions.
	 
	 Transformer-based models have shown promising potential in sequential recommendation~\cite{kang2018self,lian2020geography,sun2019bert4rec,chen2021tops,li2020time,wu2020sse,wu2020deja,liutr2021}, due to their ability of modeling arbitrary dependencies
      in a sequence. For example, SASRec~\cite{kang2018self} first adopts self-attention network to learn the importance of items at different positions. In the follow-up studies, several Transformer variants have been designed for different scenarios by adding  bidirectional attentions~\cite{sun2019bert4rec}, time intervals~\cite{li2020time}, personalization~\cite{wu2020sse}, importance sampling~\cite{lian2020geography}, and sequence augmentation~\cite{liutr2021}. However, very few studies pay attention to the robustness of self-attentive recommender models. Typically, users' sequences contain lots of irrelevant items since they may subsequently purchase a series of products with  different purposes~\cite{wang2021denoising}. As such, the current user action only depends on a subset of items, not on the  entire sequences.   However, the self-attention module is known to be sensitive to noisy sequences~\cite{kim2021lipschitz}, which may lead to sub-optimal generalization performance.  In this paper,  we aim to reap the benefits of Transformers while denoising the noisy item sequences by using learnable masks in an end-to-end fashion.

\subsection{Sparse Transformer}
Recently, many lightweight Transformers seek to achieve sparse attention maps since not all attentions carry important information in the self-attention layers~\cite{guo2019star,child2019generating,beltagy2020longformer,zaheer2020bigbird,kitaev2019reformer}. For instance,  Reformer~\cite{kitaev2019reformer} computes attentions based on  locality-sensitive hashing, leading to lower memory consumption.  Star Transformer~\cite{guo2019star} replaces the fully-connected structure of self-attention with a star-shape topology. Sparse Transformer~\cite{child2019generating} and Longformer~\cite{beltagy2020longformer} achieve sparsity by using various sparse patterns, such as diagonal sliding windows, dilated sliding windows,  local and global sliding windows. BigBird~\cite{zaheer2020bigbird} uses random and several fixed patterns to build sparse blocks. It has been shown that these sparse attentions can obtain the state-of-the-art performance and greatly reduce computational complexity. However, many of them rely on fixed attention schemas that lack flexibility and require tremendous engineering efforts.

Another line of work is to use learnable attention distributions~\cite{correia2019adaptively, sukhbaatar2019adaptive,peters2019sparse,malaviya2018sparse}. Mostly, they calculate attention weights with variants of \textsl{sparsemax} that replaces the \textsl{softmax} normalization in the self-attention networks. This allows to produce both sparse and bounded attentions, yielding a compact and interpretable set of alignments. Our Rec-denoiser is related to this line of work. Instead of using  \textsl{sparsemax}, we design a trainable binary mask for the self-attention network. 
As a result, our proposed  Rec-denoiser can automatically determine which  self-attention connections should be deleted or kept in a data-driven way.

\section{Problem and Background}

 In this section, we first formulate the problem of sequential recommendation, and then revisit several   self-attentive models. We further discuss the limitations of the existing work.

\subsection{Problem Setup}
 In sequential recommendation,  let $\mathcal{U}$ be  a set of users, $\mathcal{I}$  a set of items, and $\mathcal{S}=\{\mathcal{S}^{1}, \mathcal{S}^{2}, \ldots, \mathcal{S}^{|\mathcal{U}|}\}$  a set of users' actions. We use $\mathcal{S}^u=(S_1^{u}, S_2^{u}, \ldots, S_{|S^u|}^{u})$ to denote a sequence of items for user $u \in \mathcal{U}$  in a chronological order, where $S_t^{u} \in \mathcal{I}$ is the item that user $u$ has interacted with at time  $t$, and $|S^u|$ is the length of sequence.
 
Given the interaction history $\mathcal{S}^u$, sequential recommendation seeks to predict the next item $S_{|S^u+1|}^{u}$ at time step $|S^u+1|$. During the training process ~\cite{kang2018self, sun2019bert4rec}, it will be convenient to regard the model's input as $(S_1^{u}, S_2^{u}, \ldots, S_{|S^u-1|}^{u})$ and its expected output is a shifted version of the input sequence: $(S_2^{u}, S_3^{u}, \ldots, S_{|S^u|}^{u})$.

\subsection{Self-attentive Recommenders}
 Owing to the  ability of learning long sequences, 
Transformer architectures~\cite{vaswani2017attention} have been widely used in sequential recommendation, like SASRec~\cite{kang2018self}, BERT4Rec~\cite{sun2019bert4rec}, and  TiSASRec~\cite{li2020time}. Here
we briefly introduce the design of SASRec and discuss its limitations.

\subsubsection{\textbf{Embedding Layer}}

Transformer-based recommenders maintain an item embedding table $\mathbf{T} \in \mathbb{R}^{|\mathcal{I}| \times d}$, where $d$ is the size of the embedding.  For each sequence $(S_1^{u}, S_2^{u}, \ldots, S_{|S^u-1|}^{u})$, it can be converted into a fixed-length sequence $(s_1, s_2, \ldots, s_n )$, where $n$ is the maximum length (\textsl{e.g.},  keeping the most recent $n$ items by truncating or padding items). The embedding for  $(s_1, s_2, \ldots, s_n )$ is denoted as $\mathbf{E} \in \mathbb{R}^{n\times d}$, which can be retrieved from the table $\mathbf{T}$. To capture the impacts of different positions, one can  inject a learnable positional embedding $\mathbf{P} \in \mathbb{R}^{n\times d}$ into the  input embedding as:
\begin{equation}
\label{eq1}
    \hat{\mathbf{E}} = \mathbf{E} + \mathbf{P},
\end{equation}
where $\hat{\mathbf{E}} \in \mathbb{R}^{n\times d}$ is an order-aware embedding, which can be directly fed to any Transformer-based models.

\subsubsection{\textbf{Transformer Block}}
A Transformer block  consists of a self-attention layer and a point-wise feed-forward layer.

\noindent \textbf{Self-attention Layer}: The key component of Transformer block is the self-attention layer that is highly efficient to uncover sequential dependencies in a sequence~\cite{vaswani2017attention}. The scaled dot-product attention is a popular attention kernel:
\begin{equation}
\label{eq2} \textsf{Attention}(\mathbf{Q}, \mathbf{K}, \mathbf{V})=\textsf{softmax}\left(\frac{\mathbf{Q K}^{T}}{\sqrt{d}}\right) \mathbf{V},
\end{equation}
where  $\textsf{Attention}(\mathbf{Q}, \mathbf{K}, \mathbf{V})  \in \mathbb{R}^{n\times d}$ is the output item representations; $\mathbf{Q}= \hat{\mathbf{E}} \mathbf{W}^Q$, $\mathbf{K}= \hat{\mathbf{E}} \mathbf{W}^K$, and $\mathbf{V}= \hat{\mathbf{E}} \mathbf{W}^V$ are the queries, keys and values, respectively; $\{\mathbf{W}^Q, \mathbf{W}^K, \mathbf{W}^V\} \in \mathbb{R}^{d \times d}$ are three projection matrices; and $\sqrt{d}$  is the scale factor to produce a softer attention distribution.   In sequential recommendation, one can utilize either  left-to-right unidirectional attentions\footnote{This can be achieved by forbidding all links between $\mathbf{Q}_i$ and $\mathbf{K}_j$ for all $j > i$.} (\textsl{e.g.}, SASRec~\cite{kang2018self} and TiSASRec~\cite{li2020time}) or  bidirectional attentions (\textsl{e.g.}, BERT4Rec~\cite{sun2019bert4rec}) to predict the next item. Moreover, one can apply $H$ attention functions in parallel to enhance expressiveness: $\mathbf{H} \gets \textsf{MultiHead}(\hat{\mathbf{E}})$~\cite{vaswani2017attention}.

\noindent \textbf{Point-wise Feed-forward Layer:}
As the self attention layer is built on linear projections, we can endow the non-linearity by introducing a point-wise feed-forward layers: 
\begin{equation}
\label{eq5}
\begin{aligned}
\mathbf{F_i} = \textsf{FFN}(\mathbf{H}_i) = \textsf{ReLU}(\mathbf{H}_i \mathbf{W}^{(1)} + \mathbf{b}^{(1)})\mathbf{W}^{(2)} + \mathbf{b}^{(2)},
\end{aligned}
\end{equation}
where $\mathbf{W}^{(*)} \in \mathbb{R}^{d \times d}$,  $\mathbf{b}^{(*)} \in \mathbb{R}^{d}$ are the learnable weights and bias.

In practice,  it is usually beneficial to learn  hierarchical item dependencies by stacking more Transformer blocks. Also, one can adopt the tricks of residual connection, dropout, and layer normalization for stabilizing and accelerating the training. Here we simply summarize the output of $L$ Transformer blocks as: $\mathbf{F}^{(L)} \gets \text{Transformer} (\hat{\mathbf{E}})$.

\subsubsection{\textbf{Learning Objective}}
After stacked $L$ Transformer blocks, one can predict the next item (given the first $t$ items) based on $\mathbf{F}_t^{(L)}$. In this work, we use inner product to predict the relevance of item $i$ as:
\begin{equation*}
\label{eq7}
r_{i,t} = \langle \mathbf{F}_t^{(L)}, \mathbf{T}_i \rangle,
\end{equation*}
where $\mathbf{T}_i \in \mathbb{R}^d$ is the embedding of item $i$. Recall that the model inputs a sequence  $\mathbf{s} = (s_1, s_2, \ldots, s_n )$ and its desired output is a shifted version of the same sequence $\mathbf{o} = (o_1, o_2, \ldots, o_n )$, we can adopt the binary cross-entropy loss as:
\begin{equation}
\label{eq7}
\begin{aligned}
\mathcal{L}_{BCE}= -\sum_{\mathcal{S}^{u} \in \mathcal{S}} \sum_{t=1}^n\left[\log  (\sigma(r_{o_{t}, t}))+  \log (1-\sigma(r_{o'_{t}, t}))\right] + \alpha \cdot \|\mathbf{\Theta}\|_F^2,
\end{aligned}
\end{equation}
where $\mathbf{\Theta}$ is the model parameters, $\alpha$ is the regularizer to prevent over-fitting,  $o'_{t} \not\in \mathcal{S}^{u}$ is a negative sample corresponding to $o_{t}$, and $\sigma(\cdot)$ is the sigmoid function. More details   can be found in SASRec~\cite{kang2018self} and BERT4Rec~\cite{sun2019bert4rec}.

\subsection{The Noisy Attentions Problem}
Despite  the success of SASRec and its variants, we argue that they are insufficient to address the noisy items in sequences. The reason is that the full attention distributions (\textsl{e.g.}, Eq. (\ref{eq2})) are dense and  would assign certain credits to irrelevant items. This complicates the item-item dependencies, increases the training difficulty, and even degrades the model performance. To address this issue, several attempts have been proposed to manually define sparse attention schemas in language modeling tasks~\cite{guo2019star,child2019generating,beltagy2020longformer,zaheer2020bigbird}. However, these fixed sparse  attentions cannot adapt to the input data~\cite{correia2019adaptively}, leading to sub-optimal performance.

On the other hand, several dropout techniques are specifically designed for Transformers to keep only a small portion of attentions, including LayerDrop~\cite{Fan2020Reducing}, DropHead~\cite{zhou2020scheduled}, and UniDrop~\cite{wu2021unidrop}. Nevertheless, these randomly dropout approaches are susceptible to  bias: the fact that attentions can be dropped randomly does not mean that the model allows them to be dropped, which may lead to over-aggressive pruning issues.  In contrast,  we propose a simple yet effective data-driven method  to mask out irrelevant attentions  by using differentiable masks.

\section{Rec-Denoiser}

In this section, we present our Rec-Denoiser that consists of two parts: differentiable masks for self-attention layers and Jacobian regularization for Transformer blocks.

\subsection{Differentiable Masks}
	
The self-attention layer is the cornerstone of Transformers to capture long-range dependencies. As shown in Eq. (\ref{eq2}),  the  softmax operator assigns a non-zero weight to every item. However, full attention distributions may not always be advantageous since they may cause irrelevant dependencies, unnecessary computation, and unexpected explanation.  We next put forward differentiable masks to address this concern.

	\subsubsection{\textbf{Learnable Sparse Attentions}}  Not every item in a sequence is aligned well with user preferences in the same sense that not all attentions are strictly needed in self-attention layers. Therefore, we attach each self-attention layer  with a trainable binary mask to prune noisy or task-irrelevant attentions.   Formally, for the $l$-th self-attention layer in  Eq. (\ref{eq2}), we introduce a binary matrix $\mathbf{Z}^{(l)} \in \{0, 1\}^{n \times n} $, where $\mathbf{Z}^{(l)}_{u,v}$ denotes whether the connection between query $u$ and key $v$ is present. As such, the $l$-th self-attention layer becomes:
\begin{equation}
\label{eq8}
\begin{gathered}
\mathbf{A}^{(l)} = \textsf{ softmax}\left(\frac{\mathbf{Q}^{(l)} {\mathbf{K}^{(l)}}^{T}}{\sqrt{d}}\right), \\
\mathbf{M}^{(l)} = \mathbf{A}^{(l)}   \odot \mathbf{Z}^{(l)}, \\
\textsf{Attention}(\mathbf{Q}^{(l)}, \mathbf{K}^{(l)}, \mathbf{V}^{(l)}) = \mathbf{M}^{(l)} \mathbf{V}^{(l)},
\end{gathered}
\end{equation}
where $\mathbf{A}^{(l)}$ is the original full attentions, $\mathbf{M}^{(l)}$ denotes the sparse attentions, and  $\odot$ is the element-wise product. Intuitively, the mask $\mathbf{Z}^{(l)}$  (\textsl{e.g.}, $1$ is kept and $0$ is dropped) requires minimal changes to the original self-attention layer. More importantly, they are   capable of yielding exactly zero attention scores for irrelevant dependencies, resulting in better interpretability.   The idea of differentiable masks is not new. In the language modeling, differentiable masks have been shown to be very powerful to extract  short yet sufficient sentences, which achieves better performance~\cite{bastings2019interpretable,de2020decisions}. 

One way to  encourage sparsity of $\mathbf{M}^{(l)}$   is to   explicitly penalize the number of non-zero entries of $\mathbf{Z}^{(l)}$, for $1 \le l \le L$, by minimizing:
	\begin{equation}
	\label{eq9}
	\mathcal{R}_M = \sum_{l=1}^{L}\left\|\mathbf{Z}^{(l)}\right\|_{0}=   	\sum_{l=1}^{L} 	\sum_{u=1}^{n} 	\sum_{v=1}^{n}   \mathbb{I}\left[\mathbf{Z}^{(l)}_{u, v} \neq 0\right],
	\end{equation}
	where $\mathbb{I}[c]$ is an indicator  that is equal to $1$ if the condition $c$ holds and $0$ otherwise;  and $\|\cdot\|_0$ denotes the $L_0$ norm that is able to drive irrelevant attentions to be exact zeros. 
	
	However, there are two challenges for optimizing $\mathbf{Z}^{(l)}$: non-differentiability and large variance. $L_0$ is discontinuous  and has zero derivatives almost everywhere. Additionally, there are $2^{n^2}$ possible states for the binary mask $\mathbf{Z}^{(l)}$ with large variance. Next,  we propose an efficient estimator to solve this stochastic binary optimization problem.
	
	\subsubsection{\textbf{Efficient Gradient Computation}} Since  $\mathbf{Z}^{(l)}$ is jointly optimized with the original Transformer-based models, we combine Eq. (\ref{eq7}) and Eq. (\ref{eq9}) into one unified objective:
	\begin{equation}
	\label{eq10}
	\mathcal{L} (\mathbf{Z}, \mathbf{\Theta})= \mathcal{L}_{BCE}(\{\mathbf{A}^{(l)} \odot \mathbf{Z}^{(l)}\}, \mathbf{\Theta})+\beta  \cdot  	\sum_{l=1}^{L} 	\sum_{u=1}^{n} 	\sum_{v=1}^{n}   \mathbb{I}\left[\mathbf{Z}^{(l)}_{u, v} \neq 0\right],
	\end{equation}
	where $\beta$ controls the sparsity of masks and we denote $\mathbf{Z}$ as $\mathbf{Z}:=\{\mathbf{Z}^{(1)}, \cdots, \mathbf{Z}^{(L)}\}$. We further consider each $\mathbf{Z}^{(l)}_{u, v}$ is drawn from a Bernoulli distribution parameterized by $\mathbf{\Pi}^{(l)}_{u, v}$ such that $\mathbf{Z}^{(l)}_{u, v} \sim \text{Bern}(\mathbf{\Pi}^{(l)}_{u, v})$~\cite{louizos2017learning}.  As the parameter $\mathbf{\Pi}^{(l)}_{u, v}$ is jointly trained with the downstream tasks, a small value of $\mathbf{\Pi}^{(l)}_{u, v}$ suggests that the attention $\mathbf{A}^{(l)}_{u, v}$ is more likely to be irrelevant, and could be removed without side effects. By doing this,
	 Eq. (\ref{eq10}) becomes:
	 \begin{equation}
	 \label{eq11}
	\begin{aligned}
	{\mathcal{L}}(\mathbf{Z}, \mathbf{\Theta}) &= \mathbb{E}_{\mathbf{Z} \sim \prod^L_{l=1} \text{Bern}(\mathbf{Z}^{(l)}; \mathbf{\Pi}^{(l)})} \left[\mathcal{L}_{BCE}( \mathbf{Z}, \mathbf{\Theta})\right] +\beta   \cdot  	\sum_{l=1}^{L} 	\sum_{u=1}^{n} 	\sum_{v=1}^{n}  \mathbf{\Pi}^{(l)}_{u, v},
	\end{aligned}
	\end{equation}
	where $\mathbb{E}(\cdot)$  is the expectation. The regularization term is now continuous, but the first term $\mathcal{L}_{BCE}( \mathbf{Z}, \mathbf{\Theta})$ still involves the discrete variables $\mathbf{Z}^{(l)}$. One can address this issue by using existing gradient estimators, such as REINFORCE~\cite{williams1992simple} and Straight Through
Estimator~\cite{bengio2013estimating}, etc. These approaches, however, suffer from either biased gradients or high variance. Alternatively, we directly optimize  discrete variables using the recently proposed augment-REINFORCE-merge (ARM)~\cite{arms2021,yin2019arm,dong2020disarm},  which is unbiased and has low variance.

    In particular, we adopt the reparameterization trick~\cite{jang2016categorical}, which reparameterizes $\mathbf{\Pi}^{(l)}_{u, v}\in [0,1]$ to a deterministic function $g(\cdot)$ with parameter $\mathbf{\Phi}^{(l)}_{u, v}$, such that:
	\begin{equation}
	\label{eq01}
	\mathbf{\Pi}^{(l)}_{u, v} = g(\mathbf{\Phi}^{(l)}_{u, v}),
	\end{equation}
	since the deterministic function $g(\cdot)$  should be bounded within $[0,1]$, we simply choose the standard sigmoid function as our  deterministic function: $g(x) = {1} /( {1+e^{-x}})$. As such, the second term in Eq. (\ref{eq11}) becomes differentiable with the continuous function $g(\cdot)$. We next present the ARM estimator for computing the gradients of  binary variables in the first term of  Eq. (\ref{eq11})~\cite{arms2021,yin2019arm,dong2020disarm}.

	According to Theorem 1 in ARM~\cite{yin2019arm}, we can compute the gradients for  Eq. (\ref{eq11}) as:
	\begin{equation}
	\label{eq12}
	\begin{aligned}
	\nabla^{ARM}_{\mathbf{\Phi}} {\mathcal{L}}(\mathbf{\Phi}, \mathbf{\Theta}) &=  \mathbb{E}_{\mathbf{U} \sim \prod_{l=1}^{L} \operatorname{Uni}( \mathbf{U}^{(l)} ;0,1)} [  ( \mathcal{L}_{BCE}(\mathbb{I}[\mathbf{U}>g(-\mathbf{\Phi})], \mathbf{\Theta})-  \mathcal{L}_{BCE}(\mathbb{I}[\mathbf{U}<g(\mathbf{\Phi})], \mathbf{\Theta})) \cdot (\mathbf{U}- \frac{1}{2}) ]+ \beta \cdot \nabla_{\mathbf{\Phi}}g(\mathbf{\Phi}),
	\end{aligned}
	\end{equation}
   where $\text{Uni}(0,1)$ denotes the Uniform distribution within $[0,1]$, and $\mathcal{L}_{BCE}(\mathbb{I}[\mathbf{U}>g(-\mathbf{\Phi})], \mathbf{\Theta})$ is the cross-entropy loss obtained by setting the binary masks $\mathbf{Z}^{(l)}$ to $1$ if $\mathbf{U}^{(l)}>g(-\mathbf{\Phi}^{(l)})$ in the forward pass, and  $0$ otherwise. The same strategy is applied to $\mathcal{L}_{BCE}(\mathbb{I}[\mathbf{U}<g(\mathbf{\Phi})], \mathbf{\Theta})$.  Moreover, ARM is an unbiased estimator due to the linearity of expectations. 
   
   Note  that we need to evaluate $\mathcal{L}_{BCE}(\cdot)$ twice to compute gradients in  Eq. (\ref{eq12}). Given the fact that $u \sim \text{Uni}(0, 1)$ implies $(1-u) \sim \text{Uni}(0, 1)$,  we can  replace $\mathbf{U}$  with  $(1- \mathbf{U})$ in the indicator function inside  $\mathcal{L}_{BCE}(\mathbb{I}[\mathbf{U}>g(-\mathbf{\Phi})], \mathbf{\Theta})$:
   \begin{equation*}
    \mathbb{I}[\mathbf{U}>g(-\mathbf{\Phi})] \Leftrightarrow  \mathbb{I}[1 - \mathbf{U}>g(-\mathbf{\Phi})]  \Leftrightarrow   \mathbb{I}[\mathbf{U}<g(\mathbf{\Phi})].
   \end{equation*}
   
   To this end, we can further reduce the complexity by   considering the variants of ARM -- Augment-Reinforce (AR)~\cite{yin2019arm}:
   	\begin{equation}
	\label{eq13}
	\begin{aligned}
	&\nabla^{AR}_{\mathbf{\Phi}} {\mathcal{L}}(\mathbf{\Phi}, \mathbf{\Theta})=\mathbb{E}_{\mathbf{U} \sim \prod_{l=1}^{L} \operatorname{Uni}( \mathbf{U}^{(l)} ;0,1)} \left[    \mathcal{L}_{BCE}(\mathbb{I}[\mathbf{U}<g(\mathbf{\Phi})], \mathbf{\Theta}) \cdot (1-2\mathbf{U})  \right]  +  \beta \cdot \nabla_{\mathbf{\Phi}}g(\mathbf{\Phi}),
	\end{aligned}
	\end{equation}
	where only requires one-forward pass. The  gradient estimator $\nabla^{AR}_{\mathbf{\Phi}} {\mathcal{L}}(\mathbf{\Phi}, \mathbf{\Theta})$ is still unbiased but may pose higher variance, comparing to $\nabla^{ARM}_{\mathbf{\Phi}} {\mathcal{L}}(\mathbf{\Phi}, \mathbf{\Theta})$. In the experiments, we can trade off the variance of the estimator with complexity.
	
	In the training stage, we  update $\nabla_{\mathbf{\Phi}} {\mathcal{L}}(\mathbf{\Phi}, \mathbf{\Theta})$ (either Eq. (\ref{eq12}) or Eq. (\ref{eq13})) and $\nabla_{\mathbf{\Theta}} {\mathcal{L}}(\mathbf{\Phi}, \mathbf{\Theta})$\footnote{$\nabla_{\mathbf{\Theta}} {\mathcal{L}}(\mathbf{\Phi}, \mathbf{\Theta})$ is the original optimization for Transformers, and one can  learn $\mathbf{\Theta}$
via standard Stochastic Gradient Descent.} during the back propagation. In the inference stage, we can use the expectation of $\mathbf{Z}^{(l)}_{u, v} \sim \text{Bern}(\mathbf{\Pi}^{(l)}_{u, v})$ as the mask in Eq. (\ref{eq8}), i.e., $\mathbb{E}(\mathbf{Z}^{(l)}_{u, v}) = \mathbf{\Pi}^{(l)}_{u, v}= g(\boldsymbol{\Phi}^{(l)}_{u, v})$. Nevertheless, this will not yield a sparse attention $\mathbf{M}^{(l)}$  since the sigmoid function is smooth unless the hard sigmoid function is used  in Eq. (\ref{eq01}). Here we simply clip those values  $g(\boldsymbol{\Phi}^{(l)}_{u, v}) \le 0.5$ to zeros such that a sparse attention matrix is guaranteed and the corresponding noisy attentions are eventually eliminated.

\subsection{Jacobian Regularization} As recently proved by ~\cite{kim2021lipschitz}, the standard dot-product self-attention  is \textsl{not}  Lipschitz continuous and is vulnerable to the quality of input sequences. Let $f^{(l)}$ be the $l$-th Transformer block (Sec 3.2.2) that contains both a self-attention layer and a point-wise feed-forward layer, and $\mathbf{x}$ be the input. We can measure the robustness of the Transformer block using the residual error: $f^{(l)} (\mathbf{x} + \boldsymbol{\epsilon}) - f^{(l)} (\mathbf{x})$, where $\boldsymbol{\epsilon}$ is a small perturbation vector and the norm of  $\boldsymbol{\epsilon}$ is bounded by a small scalar $\delta$, \textsl{i.e.}, $\|\boldsymbol{\epsilon}\|_2 \le \delta$. Following the Taylor expansion, we have:
	\begin{equation*}
f^{(l)}_{i}(\mathbf{x}+\boldsymbol{\epsilon})-f^{(l)}_{i}(\mathbf{x}) \approx\left[\frac{\partial f^{(l)}_{i}(\mathbf{x})}{\partial \mathbf{x}}\right]^{\top} \boldsymbol{\epsilon}.
\end{equation*}
Let $\mathbf{J}^{(l)}(\mathbf{x})$ represent the Jacobian matrix at $\mathbf{x}$ where $\mathbf{J}^{(l)}_{ij}=\frac{\partial f^{(l)}_{i}(\mathbf{x})} { \partial x_{j} }$. Then, we  set $\mathbf{J}^{(l)}_{i}(\mathbf{x}) = \frac{\partial f^{(l)}_{i}(\mathbf{x})} { \partial \mathbf{x}}$ to denote the $i$-th row of $\mathbf{J}^{(l)}(\mathbf{x}) .$ According to Hölder's inequality\footnote{https://en.wikipedia.org/wiki/Hölder's\_inequality}, we have:
	\begin{equation*}
\|f^{(l)}_{i}(\mathbf{x}+\boldsymbol{\epsilon})-f^{(l)}_{i}(\mathbf{x}) \|_2 \approx \| \mathbf{J}^{(l)}_{i}(\mathbf{x}) ^{\top} \boldsymbol{\epsilon}\|_2 \le\| \mathbf{J}^{(l)}_{i}(\mathbf{x}) ^{\top}\|_2 \| \boldsymbol{\epsilon}\|_2.
\end{equation*}

Above inequality indicates that regularizing the $L_2$ norm on the Jacobians enforces a Lipschitz constraint at least \textsl{locally}, and the residual error is strictly bounded. Thus, we propose to regularize
 Jacobians with Frobenius norm for each Transformer block as:
	\begin{equation}
		\label{eq14}
\mathcal{R}_J = \sum_{l=1}^{L} \|\mathbf{J}^{(l)}\|_F^2.
\end{equation}

Importantly, $\|\mathbf{J}^{(l)}\|_F^2$ can be approximated via various Monte-Carlo estimators~\cite{hutchinson1989stochastic,meyer2021hutch++}. In this work, we adopt the classical Hutchinson
estimator~\cite{hutchinson1989stochastic}. For each Jocobian matrix $\mathbf{J}^{(l)} \in \mathbb{R}^{n \times n}$, we have:
\begin{equation*}
\|\mathbf{J}^{(l)}\|_F^2=
\operatorname{Tr}\left(\mathbf{J}^{(l)}  {\mathbf{J}^{(l)}} ^{\top}\right)=\mathbb{E}_{\boldsymbol{\eta} \in \mathcal{N}\left(0, \mathbf{I}_{n}\right)}\left[\left\|\boldsymbol{\eta}^{\top} \mathbf{J}^{(l)}\right\|_{F}^{2}\right],
\end{equation*}
where $\boldsymbol{\eta} \in \mathcal{N}(0, \mathbf{I}_{n})$ is the normal distribution vector. We further make use of random projections to compute the norm of Jacobians $\mathcal{R}_J$ and its gradient 	$\nabla_{\mathbf{\Theta}} \mathcal{R}_J(\mathbf{\Theta})$~\cite{hoffman2019robust}, which significantly reduces the running time in practice.

	\begin{algorithm}
	\small
		\DontPrintSemicolon % Some LaTeX compilers require you to use \dontprintsemicolon instead
		\KwIn{ The training sequences $\mathcal{S}$, the number of attention layers $L$, the number of heads $H$, and the regularization coefficients $\alpha$, $\beta$, and $\gamma$.}
		
		Initialize model parameters $\mathbf{\Theta}$, mask parameters  $\mathbf{\Phi}$;\;

		\For{\text{each mini-batch}}{
			\For{$l \gets 1$ \textbf{to} $L$}{
				Compute the full attention $\mathbf{A}^{(l)}$ in Eq. (\ref{eq8})\;
				Draw  $\mathbf{U}^{(l)} \sim \text{Uniform}(0, 1)$; \;
				Compute the mask $\mathbf{Z}^{(l)} = \mathbb{I}[ \mathbf{U}^{(l)}  < g(\mathbf{\Phi}^{(l)}) ]$;\;
				Compute the sparse attention $\mathbf{M}^{(l)}=\mathbf{A}^{(l)}   \odot \mathbf{Z}^{(l)}$;\;
				Feed $\mathbf{M}^{(l)}$ to the rest modules of Transformer;\;
				
			}
			Compute the loss $\mathcal{L}_{Rec-Denoiser}$ in Eq. (\ref{eqq}); \;
			Update the model parameters $\mathbf{\Theta}$ and mask parameters $\mathbf{\Phi}$ (Eq. (\ref{eq13})) via Stochastic Gradient Descent;\;
		}
		\caption{Rec-denoiser with AR estimator}
		\KwOut{ A well-trained Transformer.}
	\end{algorithm}

\subsection{Optimization}
	\subsubsection{\textbf{Joint Training}}
   Putting together loss in Eq. (\ref{eq7}), Eq. (\ref{eq9}), and Eq. (\ref{eq14}), the overall objective  of Rec-Denoiser is:
	\begin{equation}
	\label{eqq}
	\mathcal{L}_{ Rec-Denoiser} = \mathcal{L}_{BCE} + \beta \cdot \mathcal{R}_M  + \gamma \cdot \mathcal{R}_J,
	\end{equation}
	where $ \beta$ and $\gamma$ are regularizers  to control the  sparsity  and robustness of self-attention networks, respectively.  Algorithm 1 summarizes the overall training of Rec-Denoiser with the AR estimator.

Lastly, it is worth mentioning that our Rec-Denoiser is compatible to many Transformer-based sequential recommender models since our differentiable masks and gradient regularizations will not change their main architectures. If we simply set all masks $\mathbf{Z}^{(l)}$ to be all-ones matrix and $\beta=\gamma=0$, our model boils down to their original designs.   If we randomly set subset of masks $\mathbf{Z}^{(l)}$ to be zeros, it is equivalent to structured Dropout like LayerDrop~\cite{Fan2020Reducing}, DropHead~\cite{zhou2020scheduled}. In addition, our Rec-Denoiser can work together with  linearized self-attention networks~\cite{katharopoulos2020transformers,zhou2021informer} to further reduce the complexity of attentions. We leave this extension in the future.

	\subsubsection{\textbf{Model Complexity}} The complexity of Rec-Denoiser comes from three parts: a basic Transformer,  differentiable masks, and  Jacobian regularization. The complexity of basic Transformer keeps the same as SASRec~\cite{kang2018self} or BERT4Rec~\cite{sun2019bert4rec}. The complexity of differentiable masks requires either one-forward pass (\textsl{e.g.}, AR with high variance) or two-forward pass (\textsl{e.g.}, ARM with low variance) of the model. In sequential recommenders, the number of Transformer blocks is often very small (\textsl{e.g.}, $L=2$ in SASRec~\cite{kang2018self} and BERT4Rec~\cite{sun2019bert4rec} ). It is thus reasonable to use the ARM estimator without heavy computations. Besides, we compare the performance of AR and ARM estimators in Sec 5.3.
	
	Moreover, the random project techniques are surprisingly efficient to compute the norms of Jacobians~\cite{hoffman2019robust}. As a result, the overall computational complexity remains the same order as the original Transformers during the training. However, during the inference,  our attention maps are very sparse, which enables much faster feed-forward computations.

\section{Experiments}
Here we present our empirical results. Our experiments are designed to answer the following research questions:

\begin{itemize}
		\item \textbf{RQ1:} How effective is the proposed Rec-Denoiser compared to  the state-of-the-art sequential recommenders? 
		\item \textbf{RQ2:} How can  Rec-Denoiser reduce the negative impacts of noisy items in  a sequence?
		\item \textbf{RQ3:} How do different components  (\textsl{e.g.}, differentiable masks and Jacobian regularization) affect the overall performance of Rec-Denoiser?
\end{itemize}

\subsection{Experimental Setting}
\subsubsection{\textbf{Dataset}}
We evaluate our models on five benchmark datasets:  \textbf{Movielens}\footnote{https://grouplens.org/datasets/movielens/20m/}, \textbf{Amazon}\footnote{https://jmcauley.ucsd.edu/data/amazon/} (we choose the three commonly used categories:  \textbf{Beauty},  \textbf{Games}, and  \textbf{Movies\&TV}), and \textbf{Steam}\footnote{https://cseweb.ucsd.edu/\textasciitilde jmcauley/datasets.html}~\cite{li2020time}. Their statistics are shown in Table \ref{tab11}. Among different datasets, MovieLens is the most
dense one while Beauty has the fewest actions per user.
We use the same procedure as ~\cite{kang2018self,rendle2010factorizing,li2020time} to perform   preprocessing and  split data into train/valid/test sets, \text{i.e.}, the last item of each user’s sequence for testing, the second-to-last for validation, and the remaining items for training.

\begin{table}
% \small
\caption{The statistics of five benchmark datasets.}
\label{tab11}
\begin{tabular}{ccccc}\toprule[1pt]
Dataset           & \#Users & \#Items & Avg actions/user & \#Actions \\ \hline
MovieLens      & 6,040   & 3,416   & 163.5             & 0.987M    \\ 
Beauty     & 51,369  & 19,369  & 4.39              & 0.225M    \\ 
 Games       & 30,935  & 12,111  & 6.46              & 0.2M      \\ 
 Movies\&TV & 40,928  & 37,564  & 25.55             & 1.05M     \\ 
Steam             & 114,796 & 8,648   & 7.58              & 0.87M     \\ \toprule[1pt]
\end{tabular}
\end{table}

\subsubsection{\textbf{ Baselines}}
Here we include two groups of baselines. The first group includes general sequential  methods (Sec 5.2): 1) \textbf{FPMC}~\cite{rendle2010factorizing}: a  mixture of  matrix  factorization  and  first-order Markov chains model; 2) \textbf{GRU4Rec}~\cite{HidasiKBT15}: a RNN-based method  that  models  user  action  sequences;  3) \textbf{Caser}~\cite{tang2018personalized}: a CNN-based  framework that  captures  high-order relationships via convolutional operations; 4) \textbf{SASRec}~\cite{kang2018self}: a Transformer-based method that uses left-to-right self-attention layers; 5) \textbf{BERT4Rec}~\cite{sun2019bert4rec}: an architecture that is similar to SASRec, but using bidirectional self-attention layers; 6) \textbf{TiSASRec}~\cite{li2020time}: a time-aware self-attention model that further considers the relative time intervals between any two items; 7) \textbf{SSE-PT}~\cite{wu2020sse}: a framework that introduces personalization into self-attention layers; 8) \textbf{Rec-Denoiser}: our proposed Rec-Denoiser that can choose any self-attentive models as its backbone. 

The second group contains sparse Transformers (Sec 5.3): 1) \textbf{Sparse Transformer}~\cite{child2019generating}: it uses a fixed attention pattern, where only specific cells summarize previous locations   in the attention layers; 2) \textbf{$\alpha$-entmax sparse attention}~\cite{correia2019adaptively}:  it simply replaces \textsl{softmax} with $\alpha$-\textsl{entmax} to achieve sparsity.  Note that we do not compare with other popular sparse Transformers like Star Transformer~\cite{guo2019star},  Longformer~\cite{beltagy2020longformer}, and BigBird~\cite{zaheer2020bigbird}. These Transformers are  specifically designed for thousands of tokens or longer in the language modeling tasks. We leave their explorations for recommendations in the future.  We also do not compare with LayerDrop~\cite{Fan2020Reducing} and  DropHead~\cite{zhou2020scheduled} since the number of Transformer blocks and heads are often very small (\textsl{e.g.}, $L=2$ in SARRec)  in sequential recommendation.
Other sequential  architectures like memory networks~\cite{chen2018sequential,huang2018improving} and graph neural networks~\cite{wu2019session,chang2021sequential} have been outperformed by the above baselines, we simply omit these baselines and focus on Transformer-based models. The goal of experiments is to see  whether the proposed differentiable mask techniques can reduce the negative impacts of noisy items in the self-attention layers.

\subsubsection{\textbf{Evaluation metrics}} For easy  comparison, we adopt two common Top-N metrics, Hit$@N$ and
NDCG$@N$ (with default value $N=10$), to evaluate the performance of sequential models~\cite{kang2018self,li2020time,sun2019bert4rec}. Typically,  Hit$@N$ counts the rates of the ground-truth items among top-$N$ items, while NDCG$@N$ considers the position and assigns higher weights to higher positions. Following the work~\cite{kang2018self,li2020time}, for each user, we randomly sample $100$ negative items, and rank these items with the ground-truth item.
We calculate Hit$@10$ and NDCG$@10$ based on the rankings of these $101$ items.

\subsubsection{\textbf{Parameter settings}}For all baselines, we initialize the hyper-parameters as  the ones suggested by their original work. They are then well tuned on the validation set to achieve optimal performance. The  final results are conducted on the test set. We search the dimension size of items within $\{10,20,30,40,50\}$. As our Rec-Denoiser is a general plugin, we use the same hyper-parameters as the basic Transformers, \textsl{e.g.}, number of Transformer blocks, batch size, learning rate in Adam optimizer, \textsl{etc}. According to Table \ref{tab11}, we set the maximum length of item sequence $n=50$ for dense datasets MovieLens and Movies\&TV, and $n=25$ for sparse datasets Beauty, Games, and Steam. In addition, we set the number of Transformer blocks $L=2$, and the number of heads $H=2$ for self-attentive models. For Rec-Denoiser, two extra regularizers $\beta$ and $\gamma$ are both searched within  $\{10^{-1}, 10^{-2}, \ldots, 10^{-5} \}$. We choose ARM estimator due to the shallow structures of self-attentive recommenders.

\begin{table*}
% \small
\caption{Overall Performance of different models.  "RI" denotes the relative improvement of Rec-Denoisers over their backbones. The best performing results are boldfaced, and the second best ones are underlined.}
\label{t1}
\scalebox{0.80}{\begin{tabular}{c|cc|cc|cc|cc|cc}
\toprule[1pt]
Dataset           & \multicolumn{2}{c|}{MovieLens}    & \multicolumn{2}{c|}{Beauty}       & \multicolumn{2}{c|}{Games}        & \multicolumn{2}{c|}{Movies\&TV}   & \multicolumn{2}{c}{Steam}         \\
Metrics           & Hit@10          & NDCG@10         & Hit@10          & NDCG@10         & Hit@10          & NDCG@10         & Hit@10          & NDCG@10         & Hit@10          & NDCG@10         \\ \hline \hline
FPMC~\cite{rendle2010factorizing}              & 0.7478          & 0.4889          & 0.2810          & 0.1792          & 0.5231          & 0.3410          & 0.4806          & 0.3174          & 0.6012          & 0.4084          \\
GRU4Rec~\cite{HidasiKBT15}          & 0.5582          & 0.3383          & 0.2123          & 0.1205          & 0.2943          & 0.1939          & 0.4210          & 0.2343          & 0.4184          & 0.2687          \\
Caser~\cite{tang2018personalized}             & 0.7213          & 0.4672          & 0.2670          & 0.1531          & 0.4315          & 0.2652          & 0.4987          & 0.3120          & 0.7137          & 0.4810          \\ \hline
SASRec~\cite{kang2018self}            & 0.7434          & 0.5012          & 0.4345          & 0.2765          & 0.6748          & 0.4622          & 0.6521          & 0.4093          & 0.7723          & 0.5514          \\
SASRec+Denoiser   & 0.7980          & 0.5610          & 0.4783          & 0.3025          & 0.7391          & {\underline {0.5439}}    & \underline{ 0.7056}    & \underline { 0.4718}    & 0.8345          & 0.5946          \\
+RI (\%)                & 7.34\%          & 11.93\%         & 10.08\%         & 9.40\%          & 9.53\%          & 17.68\%         & 8.20\%          & 15.27\%         & 5.05\%          & 7.83\%          \\ \hline
BERT4Rec~\cite{sun2019bert4rec}          & 0.7549          & 0.5245          & 0.4528          & 0.3013          & 0.6812          & 0.4815          & 0.6701          & 0.4216          & 0.7901          & 0.5641          \\
BERT4Rec+Denoiser & \textbf{0.8045} & \textbf{0.5814} & 0.4883          & \textbf{0.3348} & \textbf{0.7415} & 0.5310          & \textbf{0.7212} & \textbf{0.4875} & \underline { 0.8410}    & \underline { 0.6223}    \\
+RI (\%)                & 6.57\%          & 10.85\%         & 7.84\%          & 11.12\%         & 8.85\%          & 10.28\%         & 7.63\%          & 15.63\%         & 6.45\%          & 10.32\%         \\ \hline
TiSASRec~\cite{li2020time}          & 0.7365          & 0.5164          & 0.4532          & 0.2911          & 0.6613          & 0.4517          & 0.6412          & 0.4034          & 0.7704          & 0.5517          \\
TiSASRec+Denoiser & 0.7954          & 0.5582          & \textbf{0.4962} & \underline { 0.3312}    & 0.7331          & 0.4984          & 0.6914          & 0.4671          & \textbf{0.8414} & \textbf{0.6320} \\
+RI (\%)                & 7.80\%          & 8.10\%          & 9.49\%          & 13.78\%         & 10.86\%         & 10.34\%         & 7.93\%          & 15.79\%         & 9.22\%          & 14.56\%         \\ \hline
SSE-PT~\cite{wu2020sse}            & 0.7413          & 0.5041          & 0.4326          & 0.2731          & 0.6810          & 0.4713          & 0.6378          & 0.4127          & 0.7641          & 0.5703          \\
SSE-PT+Denoiser   & \underline { 0.8010}    & \underline { 0.5712}    & \underline { 0.4952}    & 0.3265          & \underline { 0.7396}    & \textbf{0.5152} & 0.6972          & 0.4571          & 0.8310          & 0.6133          \\
+RI (\%)                & 8.01\%          & 13.31\%         & 14.47\%         & 19.55\%         & 8.61\%          & 11.68\%         & 9.31\%          & 10.76\%         & 8.76\%          & 13.51\%         \\ \toprule[1pt]
\end{tabular}}
\end{table*}

\subsection{Overall Performance (\textbf{RQ1})}

Table \ref{t1} presents the overall recommendation performance of all methods on the five datasets.   Our proposed Rec-denoisers  consistently obtain the best performance for all datasets. Additionally,  we  have the following observations:

\begin{itemize}[leftmargin=*]
		\item  The self-attentive sequential models (\textsl{e.g.}, SASRec, BERT4Rec, TiSASRec, and SSE-PT) generally outperform FPMC, GRU4Rec, and Caser with a large margin, verifying that the  self-attention networks have good ability of capture long-range item dependencies for the task of sequential recommendation.
		\item Comparing the original SASRec and its variants BERT4Rec, TiSASRec and SSE-PT, we find that the self-attentive models can gets benefit from incorporating additional information such as bi-directional attentions, time intervals, and user personalization.  Such auxiliary information is important to interpret the dynamic behaviors of users.
		\item The relative improvements of Rec-denoisers over their backbones are significant for all cases. For example, SASRec+Denoiser has on average $8.04\%$ improvement with respect to Hit$@10$ and over $12.42\%$ improvements with respect to NDCG$@10$.  Analogously, BERT4Rec+Denoiser outperforms the vanilla  BERT4Rec by average $7.47\%$ in Hit$@10$  and $11.64\%$ in NDCG$@10$. We also conduct the significant test between Rec-denoisers and their backbones, where all $p$-values$<1e^{-6}$, showing that the improvements of Rec-denoisers are statistically significant in all cases.
\end{itemize}

 \begin{figure*}
	\begin{center}
	\includegraphics[width=14.6cm]{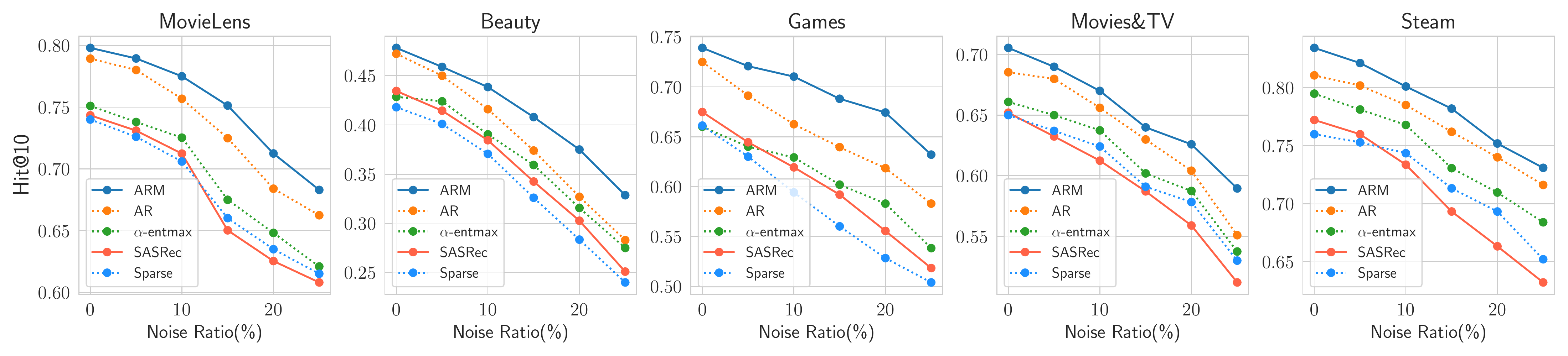}
	\end{center}
	\caption{ Overall performance on the training data are corrupted by synthetic noises.}
	\label{noise}
\end{figure*}

These improvements of our proposed models are mainly attributed to the following reasons: 1) Rec-denoisers inherit  full advantages of the self-attention networks as in  SASRec, BERT4Rec, TiSASRec, and SSE-PT; 2) Through differentiable masks, irrelevant item-item dependencies are removed, which could largely reduce the negative impacts of noisy data; 3)   Jacobian regularization enforces the smoothness of gradients, limiting  quick changes of the output against input perturbations. In general, smoothness improves the generalization of sequential recommendation. Overall, the experimental results demonstrate the superiority of
our Rec-Denoisers.

\subsection{Robustness to Noises (\textbf{RQ2})}
As discussed before,  the observed item sequences often contain some noisy items that are uncorrelated to each other. Generally, the performance of
self-attention networks is sensitive to noisy input. Here we  analyze how robust our training strategy is for noisy sequences. To achieve this, we follow the strategy~\cite{ma2020disentangled} that corrupts the training data by randomly replacing a portion of the observed items in the training set with uniformly sampled items that are not in the validation or test set.  We   range
the ratio of the corrupted training data from $0\%$ to $25\%$. We only report the results of SASRec and SASRec-Denoiser in terms of Hit$@10$. The performance of other self-attentive models is the same and omitted here  due to page limitations. In addition, we compare with two recent sparse Transformers: Sparse Transformer~\cite{child2019generating} and $\alpha$-entmax sparse attention~\cite{correia2019adaptively}.

All the simulated experiments are repeated five times and the average
results are shown in Figure \ref{noise}. Clearly, the performance of all models degrades with the increasing noise ratio.  We observe that our Rec-denoiser (use either ARM or AR estimators) consistently outperforms $\alpha$-entmax and Sparse Transformer
under different ratios of noise on all datasets. $\alpha$-entmax heavily relies on one trainable parameter $\alpha$ to filter out the noise, which may be over tuned during the training, while Sparse Transformer adopts a fixed attention pattern, which may lead to uncertain results, especially for short item sequences like Beauty and Games.

In contrast,  our differentaible masks have much more flexibility to adapt to noisy sequences. The Jacobian regularization further encourages the smoothness of our gradients, leading to better generalization.   From the results, the AR estimator performs better than $\alpha$-entmax but worse than ARM. This result is expected since ARM has much low variance.  In summary, both ARM and AR estimators are able to reduce the negative impacts of noisy sequences, which could improve the robustness of self-attentive models.

\subsection{Study of Rec-Denoiser (\textbf{RQ3})}

We further investigate the parameter sensitivity of Rec-Denoiser. For the number of blocks $L$  and the number of heads $H$, we find that self-attentive models typically benefit from small values (e.g., $H, L \leq 4$), which is similar to \cite{li2021lightweight,sun2019bert4rec}. In this section, we mainly study the following hyper-parameters:  1) the maximum length $n$, 2) the regularizers $\beta$ and $\gamma$ to control the sparsity and smoothness. Here we only study the SASRec and SASRec-Denoiser due to page limitations.

\subsubsection{\textbf{ Maximum length  $n$}} 
\begin{figure}
	\begin{center}
	\includegraphics[width=6cm]{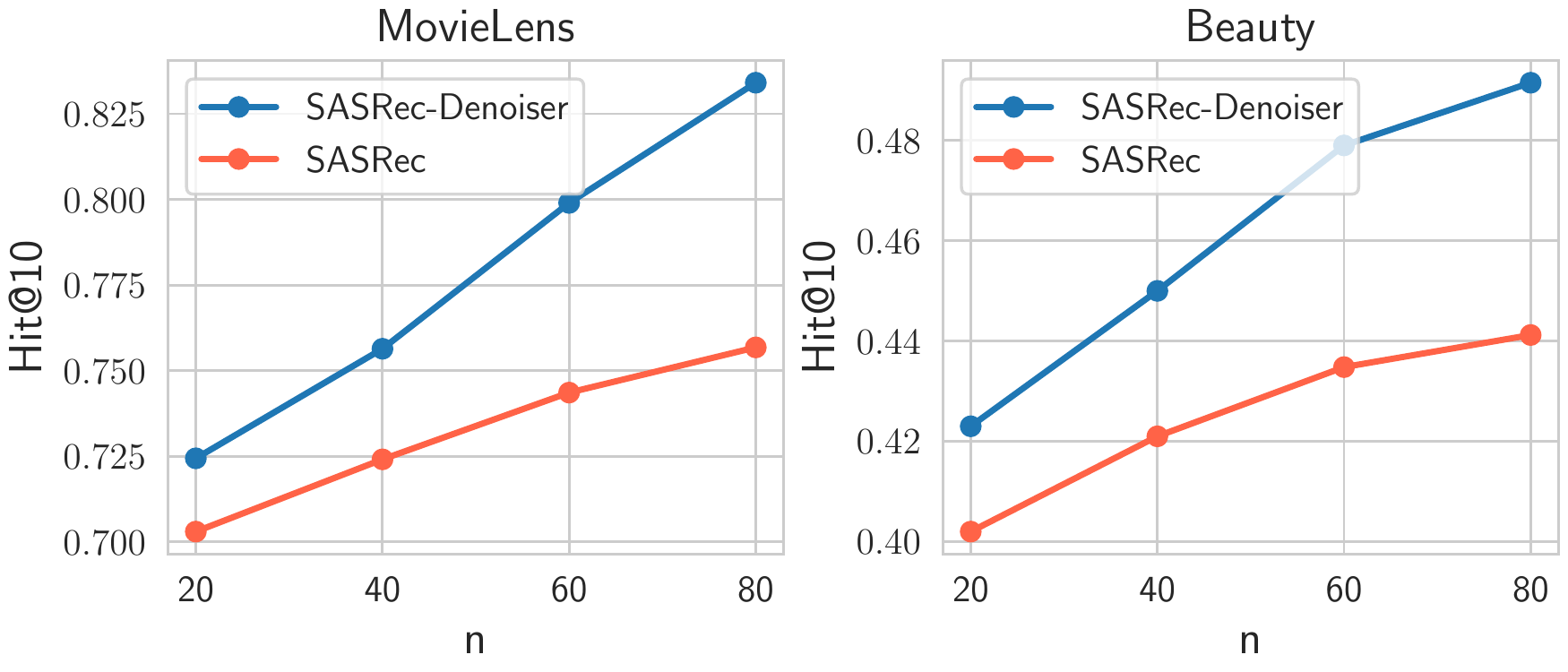}
	\end{center}
	\caption{Effect of maximum length $n$ on ranking performance (Hit$@10$).}
	\label{ran}
\end{figure}

Figure \ref{ran} shows the Hit$@10$ for maximum length $n$ from $20$ to $80$ while keeping other optimal hyper-parameters unchanged. We only test on the densest and sparsest datasets: MovieLeans and Beauty.  Intuitively, the larger sequence we have, the larger probability that the sequence contains noisy items. We observed that our SASRec-Denoiser improves the performance dramatically  with longer sequences.  This demonstrates that our design is more suitable for longer inputs, without worrying about the quality of sequences.

\subsubsection{\textbf{  The regularizers $\beta$ and $\gamma$}}
\begin{figure}
	\begin{center}
	\includegraphics[width=6cm]{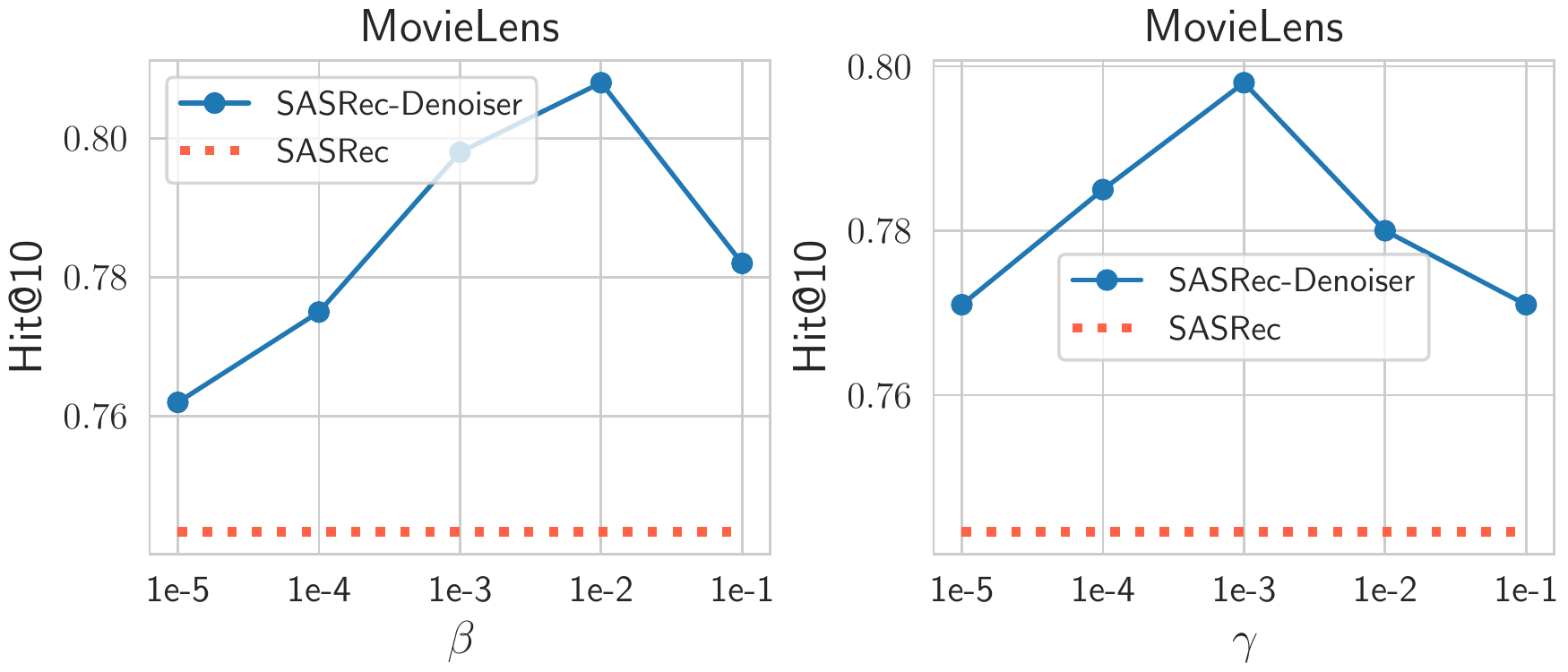}
	\end{center}
	\caption{Effect of regularizers $\beta$ and $\gamma$ on ranking performance (Hit$@10$).}
	\label{ras}
\end{figure}

There are two major regularization parameters $\beta$ and $\gamma$ for sparsity  and gradient smoothness, respectively. Figure~\ref{ras} shows the performance by changing one parameter while
fixing the other as $0.01$. As can be seen, our performance is relatively stable with respect to different settings. In the experiments, the best performance can be achieved at  $\beta=0.01$ and $\gamma=0.001$ for the MovieLens dataset. 

\section{CONCLUSION AND FUTURE WORK}
In this work, we propose Rec-Denoiser to adaptively  eliminate the negative impacts of the noisy items  for self-attentive recommender systems. The proposed Rec-Denoiser employs differentiable masks for the self-attention layers, which can dynamically prune irrelevant information.  To further tackle the vulnerability of self-attention networks to small perturbations, Jacobian regularization is applied to the Transformer blocks to improve the robustness. Our experimental results on multiple real-world sequential recommendation tasks illustrate the effectiveness of our design. 

Our proposed Rec-Denoiser framework (\textsl{e.g.}, differentiable masks and Jacobian regularization)  can be easily applied to any Transformer-based models in many tasks besides sequential recommendation. In the future, we will continue to demonstrate the contributions of our design in many real-world applications.

\bibliographystyle{ACM-Reference-Format}

\bibliography{main}

\end{document}